\documentclass[12pt]{article}
\usepackage{epsfig}
\usepackage{axodraw}
\def\bea{\begin{eqnarray}}
\def\eea{\end{eqnarray}}
\def\bq{\begin{quote}}
\def\eq{\end{quote}}

\def\gappeq{\mathrel{\rlap
{\raise.5ex\hbox{$>$}}
{\lower.5ex\hbox{$\sim$}}}}
\def\lappeq{\mathrel{\rlap{\raise.5ex\hbox{$<$}}
{\lower.5ex\hbox{$\sim$}}}}
\def\simlt{\stackrel{<}{{}_\sim}}
\def\simgt{\stackrel{>}{{}_\sim}}

\newcommand{\beq}{\begin{equation}}
\newcommand{\eeq}{\end{equation}}

\begin{document}
\pagestyle{empty}
\begin{flushright}
{\bf IFT-03/23\\
\bf hep-ph/0308032\\
\bf \today}
\end{flushright}
\vspace*{5mm}
\begin{center}
{\bf Effects of the scalar FCNC in $b\rightarrow sl^+l^-$ transitions 
and supersymmetry}
\\
\vspace*{1cm} 
Piotr H. Chankowski and \L . S\l awianowska
\\
Institute of Theoretical Physics, Warsaw University, \\
Ho\.za 69, 00-681 Warsaw, Poland 
\vskip 1.0cm

\vspace*{1.7cm} 
{\bf Abstract} 
\end{center}
\vspace*{5mm}
\noindent
{We investigate the potential effects of the scalar flavour changing neutral 
currents that are generated e.g. in supersymmetry with $\tan\beta\gg1$ in 
the $b\rightarrow sl^+l^-$ transitions. Using the experimental upper limit
on $BR(B^0_s\rightarrow\mu^+\mu^-)$ we place stringent model independent
constraints on the impact these currents may have on the rates
$BR(B\rightarrow X_s\mu^+\mu^-)$ and $BR(B\rightarrow K\mu^+\mu^-)$. 
We find that in the first case, contrary to the claim made recently in 
the literature, the maximal potential effects are always smaller than the 
uncertainty of the Standard Model NNLO prediction, that is of order 5-15\%. 
In the second case, the effects can be large but the experimental errors 
combined with the unsettled problems associated with the relevant formfactors 
do not allow for any firm conclusion about the detectability of a new physics 
signal in this process. In supersymmetry the effects of the scalar flavour 
changing neutral currents are further constrained by the experimental lower 
limit on the $B^0_s$-$\bar B^0_s$ mass difference, so that most likely no 
detectable signal of the supersymmetry generated scalar flavour changing 
neutral currents in processes $B\rightarrow X_s\mu^+\mu^-$ and 
$B\rightarrow K\mu^+\mu^-$ is possible.
}
\vspace*{1.0cm}
\date{\today} 


\vspace*{0.2cm}

\vfill\eject
\newpage

\setcounter{page}{1}
\pagestyle{plain}

\section{Introduction}

Rare processes involving the $b$-quark, intensively studied at present 
in several experiments (BaBar, BELLE, Tevatron), play an important role 
in supersymmetry (SUSY) searches via virtual effects of the new particles. 
This is because in the minimal supersymmetric extension (MSSM) of the 
Standard Model (SM) the Yukawa couplings of the $b$-quark to some of 
the superpartners of the known particles and/or to the Higgs bosons can 
be strong enough to produce measurable effects. A celebrated example is 
the radiative decay $\bar B\rightarrow X_s\gamma$ whose experimentally 
measured rate \cite{BSGexp} agrees very well with the SM prediction 
\cite{BUCZMI} and, consequently, puts constraints on the MSSM parameter
space. These constraints become particularly stringent if the ratio 
$v_u/v_d\equiv\tan\beta$ of the vacuum expectation values of the two Higgs
doublets is large, that is when the coupling of the right-chiral $b$-quark
to charginos and the top squarks is enhanced: agreement with the experimental
value can be then obtained either if all these sparticles as well as the
charged Higgs boson $H^+$ are sufficiently heavy (in which case there is
little hope to detect their virtual effects also in other rare processes), 
or if the virtual chargino-stop contribution to the $b\rightarrow s\gamma$ 
amplitude cancels against the top-charged Higgs boson contribution. The 
latter solution requires of course a certain amount of fine tuning, which 
becomes, however, of tolerable magnitude for $M_{H^+}\simgt200$ GeV and 
sparticles weighting not less than a few hundreds GeV.

A very interesting feature of the large $\tan\beta$ SUSY scenario is the 
generation at one loop of the $(\tan^2\beta)$-enhanced flavour violating
(FV) couplings of the neutral Higgs bosons, $A^0$ (the CP-odd one) and 
$H^0$ (the heavier CP-even one), to the down-type quarks \cite{CHPO1}. 
Being operators of dimension four, these couplings remain unsuppressed 
even for heavy superpartners of the known particles (gluinos, squarks, 
charginos). If the flavour violation is minimal (the so-called MFV SUSY), 
that is if the Cabbibo-Kobayashi-Maskawa (CKM) matrix is the only source 
of flavour and CP violation, the FV couplings of $A^0$ and $H^0$ are very 
sensitive to the mixing of the left and right top squarks. (Induced by 
these couplings FV decays of the neutral MSSM Higgs bosons have been 
investigated in ref. \cite{CUHETE}.) The exchanges 
of the neutral Higgs bosons generate then $|\Delta F|=1$ 
\cite{HUYA,CHGA,BAKO} and $|\Delta F|=2$ \cite{HAPOTO,BUCHROSL1} dimension 
six operators which contribute to the $b\rightarrow sl^+l^-$ and 
$b\bar s\rightarrow\bar b s$ \cite{BUCHROSL1} transitions.  For $A^0$ and 
$H^0$ not much heavier than the electroweak scale these operators, called 
because of their Lorentz structure the scalar operators, can significantly 
change the predictions of the SM.

Phenomenological consequences of the scalar operators 
have been analyzed in several papers 
\cite{HUYA,CHGA,BAKO,YAHULIZH,HULIYAZH,CHSL,BOEWKRUR,ISRE1,XIYA,BOEWKRUR2,
BUCHROSL2,CHRO,ISRE2,BUCHROSL3,DEPI,DE} both in supersymmetry with minimal 
(MFV) and nonminimal flavour violation. In particular, it has been shown 
\cite{BAKO,CHSL,BOEWKRUR,ISRE1,BOEWKRUR2} that even in the MFV SUSY the 
effects of the scalar operators originating from the FV couplings of $H^0$ 
and $A^0$ can, for large mixing of the left and right chiral top squarks, 
increase $BR(B^0_s\rightarrow\mu^+\mu^-)$ and $BR(B^0_d\rightarrow\mu^+\mu^-)$ 
by $3-4$ orders of magnitude. The upper bound on the first of these branchin 
fractions set recently by CDF \cite{NAKAO}
\begin{eqnarray}
BR(B^0_s\rightarrow\mu^+\mu^-)<0.95\times10^{-6}
\phantom{aaaa}{\rm at}\phantom{a}90\%\phantom{a}{\rm C.L.},
\label{eqn:mainlimit}
\end{eqnarray}
(which improves the previous limit 
$BR(B^0_s\rightarrow\mu^+\mu^-)<2\times10^{-6}$
\cite{CDF_Bsmm}) puts therefore on the MSSM parameter space a nontrivial 
constraint which is to a large extent complementary to the one imposed by 
the measurement of $BR(\bar B\rightarrow X_s\gamma)$. On the other hand, 
as shown in \cite{DEDRNI}, a measurement of the 
$B^0_s\rightarrow\mu^+\mu^-$ signal at the Tevatron Run II, possible if 
$BR(B^0_s\rightarrow\mu^+\mu^-)\simgt2\times10^{-8}$, would rule out such 
models of the soft SUSY breaking terms generation like anomaly and gaugino 
mediation as well as gauge mediation scenarios with low messenger scale and 
small number of messenger fields. 

The impact of the FV couplings of $H^0$ and $A^0$ on the 
$|\Delta F|=2$ transitions $B^0_s\leftrightarrow\bar B^0_s$, 
$B^0_d\leftrightarrow\bar B^0_d$ was analyzed within the MFV SUSY 
in refs. \cite{BUCHROSL1,BUCHROSL2,CHRO,BUCHROSL3}. It was found that 
the contribution of the $|\Delta F|=2$ scalar operators constructed out 
of these couplings to the amplitude of the $B^0_s$-$\bar B^0_s$ mixing 
is negative and can be very large ($B^0_d$-$\bar B^0_d$ mixing is 
affected negligibly). Part of the parameter space corresponding to 
$\tan\beta\gg1$, light $H^0$ and $A^0$ and substantial stop mixing, 
allowed by the experimental limit on $BR(B^0_s\rightarrow\mu^+\mu^-)$
then available, was eliminated by the condition that the calculated 
$B^0_s$-$\bar B^0_s$ mass difference $\Delta M_s$ is not smaller than 
the experimental lower bound $\Delta M_s\simgt14/$ps \cite{OSC_BsBs}.
Even with the new bound (\ref{eqn:mainlimit}) the constraints on the 
MSSM parameter space imposed by the $B^0_s$-$\bar B^0_s$ mixing are in 
some cases stronger than the ones stemming from the dimuon channel. 

The effects of the scalar operators in the exclusive transitions
$\bar B\rightarrow K\mu^+\mu^-$ and $\bar B\rightarrow K^\ast\mu^+\mu^-$ 
have been investigated in \cite{YAHULIZH,BOEWKRUR}. Their impact on 
$BR(\bar B\rightarrow K^\ast\mu^+\mu^-)$ has been found to be very small.
On the other hand, potential effects of the scalar operators in 
$\bar B\rightarrow K\mu^+\mu^-$ could be quite sizeable in principle, but
the experimental limit $BR(B^+\rightarrow K^+\mu^+\mu^-)<5.2\times10^{-6}$ 
\cite{AFFexp} available at that time was too weak to provide constraints 
stronger than the experimental upper limit for 
$BR(B^0_s\rightarrow\mu^+\mu^-)$. Finally, the effects of the scalar 
operators in the inclusive decay rate
$BR(\bar B\rightarrow X_s\mu^+\mu^-)$ have been taken into account in 
several papers devoted to general investigation of the potential SUSY 
effects in radiative $B$ decays or in the 
studies of the specific SUSY scenarios like the minimal SUGRA,
but have not been directly confronted with the bounds provided by the 
$B^0_s\rightarrow\mu^+\mu^-$ decay and $B^0_s$-$\bar B^0_s$ mixing.

In this paper we fill this gap. We begin in section 2 by recalling the
NNLO predictions of the SM for $BR(B^0_s\rightarrow\mu^+\mu^-)$ and
$BR(\bar B\rightarrow X_s\mu^+\mu^-)$ improving slightly in the latter 
case the estimates of the theoretical uncertainties compared to those
given in ref. \cite{ALLUGRHI}. Then in section 3, following ref. 
\cite{BOEWKRUR}, we asses in a model independent way how big effects 
of the scalar operators in the $BR(\bar B\rightarrow X_s\mu^+\mu^-)$ 
and in $BR(\bar B\rightarrow K\mu^+\mu^-)$ decays are still allowed by 
the CDF bound $BR(B^0_s\rightarrow\mu^+\mu^-)<0.95\times10^{-6}$. We show 
in particular, that the huge effects of the scalar operators found recently 
in $BR(\bar B\rightarrow X_s\mu^+\mu^-)$ in ref. \cite{WAAT} are excluded 
by these constraints. The results of section 3 are valid generally, 
independently of the mechanism that generates the scalar operators. Finally,
in section 4 we concentrate on scalar operators in the MFV version of the 
MSSM (in which the squark mass matrices are aligned with the quark ones - 
see \cite{BUCHROSL3} for more detailed explanations) and specify the maximal 
effects of the scalar operators in $BR(\bar B\rightarrow X_s\mu^+\mu^-)$ and
in $BR(\bar B\rightarrow K\mu^+\mu^-)$ allowed by the experimental limits 
on both, the $B^0_s\rightarrow\mu^+\mu^-$ rate and the $B^0_s$-$\bar B^0_s$ 
mass difference. We summarize the situation in the last section.

\section{$b\rightarrow sl^+l^-$ and  $b\rightarrow dl^+l^-$ transitions in
the SM}

Under the assumption of minimal flavour violation, the effective Hamiltonian
describing the $b\rightarrow sl^+l^-$ ($b\rightarrow dl^+l^-$) and
$b\rightarrow s\gamma$ transitions takes the form \cite{MI}
\begin{eqnarray}
{\cal H}^{\rm eff}=-2\sqrt2G_F V_{ts}^{{\rm eff}\ast} V_{tb}^{\rm eff}
\left(\sum_{X=1}^{10}C_X(\mu){\cal O}_X(\mu)
+ \sum_{l=e,\mu,\tau}\sum_{X=S,P}C^l_X(\mu){\cal O}^l_X(\mu)\right)
\label{eqn:Heff}
\end{eqnarray}
with the following set of operators ${\cal O}^{(l)}_X$ \cite{MI,BOMIUR,XIYA}
\begin{eqnarray}
&&{\cal O}_{1c}=(\bar s_L\gamma^\mu T^ac_L)
(\bar c_L\gamma_\mu T^ab_L)\nonumber\\
&&{\cal O}_{2c}=(\bar s_L\gamma^\mu c_L)
(\bar c_L\gamma_\mu b_L)\nonumber\\
&&{\cal O}_3=(\bar s_L\gamma^\mu b_L)
\sum_{q=u,d,s,c,b}(\bar q\gamma_\mu q)\nonumber\\
&&{\cal O}_4=(\bar s_L\gamma^\mu T^ab_L)
\sum_{q=u,d,s,c,b}(\bar q\gamma_\mu T^a q)\nonumber\\
&&{\cal O}_5=(\bar s_L\gamma^\mu\gamma^\nu\gamma^\lambda b_L)
\sum_{q=u,d,s,c,b}(\bar q\gamma_\mu\gamma_\nu\gamma_\lambda q)\nonumber\\
&&{\cal O}_6=(\bar s_L\gamma^\mu\gamma^\nu\gamma^\lambda T^ab_L)
\sum_{q=u,d,s,c,b}(\bar q\gamma_\mu\gamma_\nu\gamma_\lambda T^aq)\nonumber\\
&&{\cal O}_7={e\over g_s^2}(\bar s_L\sigma^{\mu\nu} b_R)F_{\mu\nu}
\label{eqn:oplist}\\
&&{\cal O}_8={1\over g_s}(\bar s_L\sigma^{\mu\nu}T^a b_R)G^a_{\mu\nu}
\nonumber\\
&&{\cal O}_9={e^2\over g_s^2}(\bar s_L\gamma^\mu b_L)
\sum_l(\bar l\gamma^\mu l)\nonumber\\
&&{\cal O}_{10}={e^2\over g_s^2}(\bar s_L\gamma^\mu b_L)
\sum_l(\bar l\gamma^\mu\gamma^5 l)\nonumber\\
&&{\cal O}_S^l={e^2\over g_s^2}(\bar s_Lb_R)
(\bar ll)\nonumber\\
&&{\cal O}_P^l={e^2\over g_s^2}(\bar s_Lb_R)
(\bar l\gamma^5l)\nonumber
\end{eqnarray}
and ${\cal O}_{1u}$, ${\cal O}_{2u}$ obtained from ${\cal O}_{1c}$ and 
${\cal O}_{2c}$ by the replacement $c\rightarrow u$, and the Wilson 
coefficients $C_X(\mu)$ organized as \cite{MI}
\begin{eqnarray}
C_X(\mu)= C^{(0)}_X(\mu)+{g^2_s(\mu)\over(4\pi)^2}C^{(1)}_X(\mu)
+{g^4_s(\mu)\over(4\pi)^4}C^{(2)}_X(\mu)+\dots\label{eqn:CXexp}
\end{eqnarray}
The coefficients $C_X$ computed at some scale $\mu_0\sim m_t$ are 
subsequently evolved down to the scale $\mu_b\sim m_b$, where their matrix 
elements between the hadronic initial and final states of the process under 
investigation are computed either by lattice methods or perturbatively to 
the required accuracy in $\alpha_s(\mu_b)=g_s^2(\mu_b)/4\pi$. At the 
matching scale $\mu_0$ only the coefficients of the operator ${\cal O}_2$ 
starts at order $(\alpha_s)^0$; for the remaining ones $C^{(0)}_X(\mu_0)=0$. 

\subsection{$B^0_{s,d}\rightarrow\mu^+\mu^-$ in the SM}

In the SM the Wilson coefficients $C_S$ and $C_P$ are negligible and the 
only operator relevant for the $B_{s,d}^0\rightarrow l^+l^-$ transitions is 
${\cal O}_{10}$. Its Wilson coefficients $C^{(1)}_{10}$ and $C^{(2)}_{10}$ 
at the matching scale are known \cite{BUBU,BOMIUR}. Since the quark part of 
${\cal O}_{10}$ is a (partially) conserved chiral current, the QCD evolution 
of $C_{10}$ is simple, i.e. $C_{10}(\mu_b)=[\alpha_s(\mu_b)/\alpha_s(\mu_0)]
C_{10}(\mu_0)$. This leads to the well known prediction \cite{BUREV}
\begin{eqnarray}
BR(B^0_q\rightarrow l^+l^-)={\tau(B_q^0)\over\pi}M_{B^0_q} 
\left({G_F\alpha_{\rm em}\hat F_{B_q}m_l\over4\pi\sin^2\theta_W}\right)^2
\sqrt{1-4{m^2_l\over M^2_{B^0_q}}}|V_{tq}^\ast V_{tb}|^2 |Y(x_t)|^2~,
\end{eqnarray}
where 
\begin{eqnarray}
{1\over\sin^2\theta_W}Y(x_t)=
C^{(1)}_{10}(x_t)+{g^2_s(\mu_0)\over16\pi^2}C^{(2)}_{10}(x_t,\mu_0)
\end{eqnarray}
and $x_t=(m_t^{\overline{\rm MS}}(\mu_0)/M_W)^2$. $C^{(1)}_{10}(x_t)$ is 
given by the function $Y_0(x_t)$, which can be found e.g. in \cite{BUREV} 
and $C^{(2)}_{10}(x_t,\mu_0)$ has been computed in \cite{BUBU} 
(it can be also extracted from \cite{BOMIUR}). For 
$m_t^{\overline{\rm MS}}(m_t)=(166\pm5)$ GeV, $\alpha_s(M_Z)=0.119$  and
using $\mu_0=m_t=174.3$ GeV
\begin{eqnarray}
Y(x_t)=\eta ~(0.971\pm0.046)
\end{eqnarray}
where $\eta=1.01$ accounts for the effects of $C^{(2)}_{10}$.
For $\sin^2\theta_W=0.23124$ and $\alpha_{\rm em}=1/128$ this gives 
\begin{eqnarray}
BR(B^0_s\rightarrow\mu^+\mu^-)=(3.64\pm0.33)\times10^{-9}\times
\left({\tau_{B^0_s}\over1.461 ~{\rm ps}}\right)
\left({\hat F_{B_s}\over238 ~{\rm MeV}}\right)^2
\left({|V_{ts}|\over0.04}\right)^2\nonumber\\
BR(B^0_d\rightarrow\mu^+\mu^-)=(1.39\pm0.13)\times10^{-10}\times
\left({\tau_{B^0_d}\over1.542 ~{\rm ps}}\right)
\left({\hat F_{B_d}\over203 ~{\rm MeV}}\right)^2
\left({|V_{td}|\over0.009}\right)^2\label{eqn:B0llSM}
\end{eqnarray}
where the errors correspond to the variation of 
$m_t^{\overline{\rm MS}}(m_t)$. The dominant uncertainities of the SM 
predictions (of order $\sim^{+28}_{-24}$\% and $\sim^{+40}_{-30}$\% in 
the case of the $B^0_s$ and $B^0_d$ decays, respectively) come from the 
factors $\hat F_{B_s}=(238\pm31)$ MeV and 
$\hat F_{B_d}=(203\pm27^{+0}_{-20})$ MeV \cite{LEBE} that parametrize the 
nonperturbative hadronic matrix element of the ${\cal O}_{10}$ operator. 
The uncertainty associated with $\Delta m_t^{\overline{\rm MS}}(m_t)=5$ 
GeV, with the electromagnetic corrections and, in the case the $B^0_d$ 
decay with the value of $|V_{td}|$, are much smaller. 

The corresponding 
branching ratios for the $e^+e^-$ channel are suppressed by the factor
$(m_e/m_\mu)^2\sim2\times10^{-5}$ and, hence, unmeasurably small; those 
for the $\tau^+\tau^-$ channel are enhanced by $(m_\tau/m_\mu)^2\sim283$
but taons are very difficult to identify experimentally.

The present experimental bounds 
$BR(B^0_s\rightarrow\mu^+\mu^-)<0.95\times10^{-6}$ \cite{NAKAO} and 
$BR(B^0_d\rightarrow\mu^+\mu^-)<1.6\times10^{-7}$ \cite{BABAR_Bdmm,NAKAO} 
are 3 orders of magnitude above the predictions (\ref{eqn:B0llSM}) and 
still leave a lot of room for new physics. 

\subsection{The inclusive process $\bar B\rightarrow X_sl^+l^-$ in the SM}

The general formula for the differential width of the 
$B\rightarrow X_sl^+l^-$ decay reads \cite{XIYA,WAAT}:
\begin{eqnarray}
{d\over ds}\Gamma(\bar B\rightarrow X_s\mu^+\mu^-)
={G^2_F\alpha^2_{\rm em}m_b^5\over768\pi^5}|V_{tq}^\ast V_{tb}|^2
\lambda^{1/2}(1,r_s,s)\lambda^{1/2}(1,r_s/s,r_s/s)\phantom{aaaaaaaaaa}
\nonumber\\
\times\left\{G_c(s)
+f_1(s)G_1(s)\left|\tilde C_9^{\rm eff}(s,\mu_b)\right|^2
+f_2(s)G_1(s)\left|\tilde C_{10}^{\rm eff}(s,\mu_b)
\right|^2\phantom{aaaaa}
\right.\label{eqn:diffwidth}\\
+f_3(s)G_2(s)\left|\tilde C_7^{\rm eff}(s,\mu_b)\right|^2
+f_4(s)G_3(s){\rm Re}
\left(\tilde C_7^{\rm eff}(s,\mu_b)\tilde C_9^{{\rm eff}\ast}(s,\mu_b)\right)
\phantom{aaa}\nonumber\\
\left.
+f_5(s)\left|C^{(1)}_S(\mu_b)\right|^2+f_6(s)\left|C^{(1)}_P(\mu_b)\right|^2
+f_7(s){\rm Re}\left(\tilde C_{10}^{\rm eff}(s,\mu_b) 
C^{(1)\ast}_P(\mu_b)\right)
\right\}
\nonumber
\end{eqnarray}
where $s=q^2/m^2_b$ is the ``reduced'' invariant mass of the lepton pair and 
\begin{eqnarray}
\lambda(a,b,c)=a^2+b^2+c^2-2ab-2ac-2bc~.
\end{eqnarray}
The function $G_c(s,\lambda_1,\lambda_2)$ accounting for the $1/m_c^2$ 
nonperturbative contribution has been found in \cite{BUISRE}. The 
$1/m_b^2$ nonperturbative contributions summarized by the functions 
$G_i(s,\lambda_1,\lambda_2)$ have been calculated using the heavy 
quark expansion technique in \cite{ALHIHAMO,BUIS}. The functions 
$G_c(s,\lambda_1,\lambda_2)$ and $G_i(s,\lambda_1,\lambda_2)$, which
depend on the parameters $\lambda_1\approx-0.2$ GeV$^2$, $\lambda_2=0.12$ 
GeV$^2$ are given in eqs. (29-31) of \cite{ALLUGRHI}. Finally,\footnote{The 
functions $f_3(s)$ and $f_7(s)$ differ from the corresponding expressions 
in ref. \cite{WAAT}. Due to the extra piece $-s^2$ the function $f_3(s)$ 
as given here reproduces in the limit $m_s=0$ the  result obtained in 
earlier papers for the coefficient of $|\tilde C_7^{\rm eff}|^2$. We also 
confirm that the sign of $f_7(s)$ is as in the earlier papers \cite{XIYA} 
(opposite to the one in \cite{WAAT}).}
\begin{eqnarray}
&&f_1(s)=s(1+r_s-s)\lambda(1,r_s/s,r_s/s)+(1-r_s+s)(1-r_s-s)(1+2r_s/s)
\nonumber\\
&&\phantom{aaaa}+6r_l(1+r_s-s)\nonumber\\
&&f_2(s)=s(1+r_s-s)\lambda(1,r_s/s,r_s/s)+(1-r_s+s)(1-r_s-s)(1+2r_s/s)
\nonumber\\
&&\phantom{aaaa}-6r_l(1+r_s-s)\nonumber\\
&&f_3(s)=(4/s)(1+2r_s/2)\left[2(1+r_s)(1-r_s)^2-s(1+14r_s+r_s^2)-s^2\right]
\\
&&f_4(s)=12(1+2r_s/s)\left[(1-r_s)^2-s(1+r_s)\right]
\nonumber\\
&&f_5(s)={3\over2}(1+r_s-s)(s-4r_l)
\nonumber\\
&&f_6(s)={3\over2}(1+r_s-s)s
\nonumber\\
&&f_7(s)=6\sqrt{r_l}(1-r_s-s)\nonumber
\end{eqnarray}
where $r_l=m^2_l/m^2_b$, $r_s=m^2_s/m^2_b$. In the NNLO approximation 
the coefficients $\tilde C_7^{\rm eff}(s,\mu_b)$, 
$\tilde C_9^{\rm eff}(s,\mu_b)$ and $\tilde C_{10}^{\rm eff}(s,\mu_b)$ 
summarizing the effects of the QCD running from the scale $\mu_0\sim m_t$ 
down to the scale $\mu_b\sim m_b$ and the matrix elements of the relevant 
operators from the list (\ref{eqn:oplist}) can be compactly written as
\cite{ASASGRWA1}:
\begin{eqnarray}
&&\tilde C_7^{\rm eff}(s,\mu_b)=
\left(1+{\alpha_s(\mu_b)\over\pi}\omega_7(s)\right)A_7\nonumber\\
&&\phantom{aaaaaa}- {\alpha_s(\mu_b)\over4\pi}\left(C^{(0)}_1F^{(7)}_1(s)
+C^{(0)}_2F^{(7)}_2(s)+C^{(1)}_8F^{(7)}_8(s)\right)\nonumber\\
&&\tilde C_9^{\rm eff}(s,\mu_b)=
\left(1+{\alpha_s(\mu_b)\over\pi}\omega_9(s)\right)
\left[A_9+T_9g(m_c^2/m_b^2,s)+U_9g(1,s)+W_9g(0,s)\right]\phantom{aa}
\label{eqn:tildeC}\\
&&\phantom{aaaaaa}- {\alpha_s(\mu_b)\over4\pi}\left(C^{(0)}_1F^{(9)}_1(s)
+C^{(0)}_2F^{(9)}_2(s)+C^{(1)}_8F^{(9)}_8(s)\right)\nonumber\\
&&\tilde C_{10}^{\rm eff}(s,\mu_b)=
\left(1+{\alpha_s(\mu_b)\over\pi}\omega_9(s)\right)
A_{10}\nonumber
\end{eqnarray}
where $A_i$, $T_9$, $U_9$, $W_9$, the function $g(z,s)$ can be found in  
\cite{BOMIUR} and the explicit formulae for the functions $F^{(i)}_j(s)$ 
and $\omega_i(s)$, valid for $s\simlt0.25$ are given in 
refs.~\cite{ASASGRWA1}.\footnote{Complete results for the matrix elements, 
valid in the entire range of $s$, have been reported in \cite{GHHUISYA} 
but are not yet publicly available.}
Wilson coefficients $C^{(0)}_1$, $C^{(0)}_2$ and $C^{(1)}_8$ can be found 
e.g. in eqs. (E.9) of \cite{GAMI}. One should also remember to expand the 
formula (\ref{eqn:diffwidth}) only up to terms of order $\alpha_s(\mu_b)$ 
and to replace $\omega_7(s)$ and  $\omega_9(s)$ by $\omega_{79}(s)$ in the 
interference term. Inclusion to $\tilde C_7^{\rm eff}$, 
$\tilde C_9^{\rm eff}$ and $\tilde C_{10}^{\rm eff}$ of the 
${\cal O}(\alpha_s(\mu_b))$ corrections to the matrix 
elements\footnote{In this analysis we neglect the contribution of the real 
gluon bremsstrahlung calculated in \cite{ASASGRWA2} which changes the result 
by $\sim1$\%.}
of the relevant operators significantly decreases the dependence of the 
final result on the renormalization scale $\mu_b$ \cite{ASASGRWA1}. Since
similar ${\cal O}(\alpha_s(\mu_b))$ corrections to the matrix elements of 
the operators ${\cal O}^l_{S,P}$ are not known at present, their 
contribution to the rates of the inclusive $B\rightarrow X_sl^+l^-$ 
processes has the uncertainty associated with choice of the scale 
$\mu_b$ larger than do the contributions of the remaining operators.

In oder to get rid of the factor $m_b^5$ in the formula 
(\ref{eqn:diffwidth}) not introducing at the same time large uncertainty 
associated with the value of the charm quark mass we follow the trick 
proposed in \cite{GAMI} and normalize the rate to width of the charmless 
semileptonic decay
\begin{eqnarray}
{dBR(\bar B\rightarrow X_s\mu^+\mu^-)\over ds}
={BR(\bar B\rightarrow X_ce\nu_e)\over C}
\phantom{aaaaaaaaaaaaaaaaaaaaaa}\nonumber\\
\times{{d\over ds}\Gamma(\bar B\rightarrow X_s\mu^+\mu^-)
\over{G_F^2m_b^5\over192\pi^3}\left|V_{cb}\right|^2
\left(1-{2\alpha_s(m_b)\over3\pi}h(0)\right)
\left(1+{\lambda_1\over2m_b^2}-{9\lambda_2\over2m_b^2}\right)}
\label{eqn:Ctrick}
\end{eqnarray}
where the function $h(z)$ is given e.g. by the formula (48) of 
\cite{BOMIUR} and the factor $C$ 
\begin{eqnarray}
C\equiv \left|{V_{ub}\over V_{cb}}\right|^2
{\Gamma(\bar B\rightarrow X_ce\nu_e)\over 
\Gamma(\bar B\rightarrow X_ue\nu_e)}\label{eqn:Cfact}
\end{eqnarray}
has been calculated in \cite{GAMI}:
$C=0.575\times(1\pm0.01_{\rm pert}\pm0.02_{\lambda_1}\pm0.02_\Delta)
=0.575\times(1\pm0.03)$. To remain conservative we will double this 
uncertainty and use $C=0.575\times(1\pm0.06)$.
The poorly known nonperturbative parameter 
$\lambda_1$ approximately cancels out between the numerator and the 
denominator. With this trick the residual dependence on $z=m_c^2/m_b^2$ 
is negligible for $m_c^2/m_b^2$ varying between 0.27 and 0.31; the 
uncertainty of the differential branching fraction arising from the 
normalization is then dominated by the $\sim\pm6$\% uncertainty 
of the factor $C$ (\ref{eqn:Cfact}). It is therefore much smaller than 
the uncertainty of order $\pm15$\% attributed to the differential 
branching fraction normalized directly to $BR(\bar B\rightarrow X_ce\nu_e)$ 
in ref. \cite{ALLUGRHI} by varying $m_c^2/m_b^2$ in the range 0.25$-$0.33.

The dominant source of uncertainty remains the dependence on $\mu_b$
which for $s<0.25$ is estimated (by changing $\mu_b$ between 2.5 GeV
and 10 GeV) to be of order $\pm7$\% \cite{ASASGRWA1}. Of comparable 
magnitude can be however also the uncertainty related to the 
electromagnetic corrections to the running (and their mixing with others) 
of the ${\cal O}_9$ and ${\cal O}_{10}$ operators, which is unknown at 
present.\footnote{This conclusion has been reached in a discussion with
M. Misiak.}
Simple estimate of this effect is obtained by varying $\alpha_{\rm em}$ 
in the formula (\ref{eqn:diffwidth}) between $1/128$ and $1/133$. This 
suggests additional $\sim8$\% uncertainty of the predicted branching 
ratio. Finally, the parametric uncertainty related to the variation of 
$m_t^{\overline{\rm MS}}(m_t)=(166\pm5)$ GeV is  of order $\pm(6-7)$\%.

The differential rate (\ref{eqn:diffwidth}) can be integrated over various
domains of $s$. The most reliable theoretical predictions are obtained for 
$0.05<s<0.25$ because for this range the nonperturbative effects associated 
with the $\bar cc$ resonances are small and the NNLO calculation is 
complete. For this region, using $m_t^{\overline{\rm MS}}(m_t)=166$ GeV, 
$m_b=4.8$ GeV, $\alpha_{\rm em}=1/128$ and 
$|V_{ts}V_{tb}^\ast/V_{cb}|=0.976$ we get:
\begin{eqnarray}
BR(\bar B\rightarrow X_s\mu^+\mu^-)_{0.05<s<0.25}
=(1.46\pm0.11\pm0.10)\times10^{-6}
\end{eqnarray}
where we have used $BR(\bar B\rightarrow X_ce\nu_e)=0.102$. The first 
uncertainty comes from the $\mu_b$ dependence and the second one from 
$\Delta m_t^{\overline{\rm MS}}(m_t)=5$ GeV. To this one has to add the 
$6$\% uncertainty from the $C$ factor and (conservatively) a $\sim8$\% 
uncertainty from the electromagnetic corrections. Adding all these 
uncertainties in quadratures we finally assign to the result the 
uncertainty of order $\pm14$\%.

Integrating the differential rate (\ref{eqn:diffwidth}) over the entire
domain\footnote{Keeping $m_s\neq0$ has numerically a very small impact 
on $d\Gamma/ds$ itself but $s_{\rm max}<1$ for the upper integration limit
partly cures the problem associated with the nonperturbative contributions
to the differential rate, which for $s\rightarrow1$ dominate in the 
expression (\ref{eqn:diffwidth}) and make it negative in the vicinity of 
$s=1$ \cite{BUIS}.}
$s_{\rm min}<s<s_{\rm max}$ where $s_{\rm min}=4m_l^2/m_b^2$,
$s_{\rm max}=(1-m_s/m_b)^2$ one obtains the so-called ``nonresonant''
branching fraction which can be compared with the experimental data
provided the contribution of the $\bar cc$ resonances is judiciously
subtracted from the latter on the experimental side. Since the NNLO 
formulae for the matrix elements given in \cite{ASASGRWA1} are valid only 
for $s<0.25$, following the prescription of ref. \cite{ALLUGRHI} we have 
used for the region $s>0.25$ only the formulae of ref. \cite{BOMIUR} with 
$\mu_b=2.5$ GeV (because for $s<0.25$ the formulae of \cite{BOMIUR} with 
$\mu_b=2.5$ GeV quite accurately reproduce the full NNLO results obtained 
with $\mu_b=5$ GeV) and assigned to the integral over this range of $s$ 
the same $\mu_b$ uncertainty as has 
$d\Gamma(\bar B\rightarrow X_sl^+l^-)/ds$ computed for $s=0.25$. We get in 
this way 
\begin{eqnarray}
BR(\bar B\rightarrow X_s\mu^+\mu^-)_{\rm nonres}
=(4.39^{+0.24}_{-0.36}\pm0.24)\times10^{-6}~,\label{eqn:BRmmnonres}\\
BR(\bar B\rightarrow X_se^+e^-)_{\rm nonres}
=(7.26^{+0.25}_{-0.58}\pm0.28)\times10^{-6}~,\label{eqn:BReenonres}
\end{eqnarray}
where the meaning of the errors is as previously. Taking into account 
the remaining uncertainties we estimate the total uncertainty of 
$BR(\bar B\rightarrow X_s\mu^+\mu^-)_{\rm nonres}$ for 
$^{+13}_{-14}$\% and of $BR(\bar B\rightarrow X_se^+e^-)_{\rm nonres}$
for $^{+11}_{-14}$\%. Our central values are in good agreement 
with the ones given in ref. \cite{ALLUGRHI} but due to the normalization
to the width of the semileptonic charmless decay the overall uncertainty 
is smaller even though we take into account the uncertainties related to 
the electromagnetic correction. Within the errors and uncertainties the 
SM prediction (\ref{eqn:BRmmnonres}) is roughly in agreement with the
published BELLE \cite{BXsLLexp} and recent BaBar results, which together
give \cite{NAKAO}
$BR(\bar B\rightarrow X_sl^+l^-)_{\rm nonres}=(6.2\pm1.7)\times10^{-6}$, 
averaged over $l=\mu,e$, for the dilepton invariant mass $\sqrt{q^2}>0.2$ GeV. 
\footnote{Our result for $BR(\bar B\rightarrow X_se^+e^-)_{\rm nonres}$
for $\sqrt{q^2}>0.2$ GeV is similar to (\ref{eqn:BRmmnonres}):
$(4.48^{+0.24}_{-0.37}\pm0.24)\times10^{-6}$. In the comparison with the 
BELLE result one has to take also into account the error in translating 
the ``reduced'' invariant mass $s=q^2/m_b^2$ into the experimental cut on 
physical $q^2$.}

For a relatively clean comparison of the experimental measurements with 
the theoretical predictions of interest can be also the rate integrated 
over the region of $s$ above the $\bar cc$ resonances. We get there
\begin{eqnarray}
BR(\bar B\rightarrow X_sl^+l^-)_{0.65<s<s_{\rm max}}
=(2.32^{+0.17}_{-0.20}\pm0.14)\times10^{-7}\label{eqn:BRllhigh}
\end{eqnarray}
($l=e$ or $\mu$) where the first uncertainty, corresponding to the $\mu_b$
dependence, is estimated with the help of the prescription of ref.
\cite{ALLUGRHI} described above. Better estimate of this uncertainty will 
become possible once the calculation of ref. \cite{GHHUISYA} is available. 
However for this range of $s$ the nonperturbative $1/m_b^2$ corrections 
of refs. \cite{ALHIHAMO,BUIS} constitute yet another potential source of 
uncertainty. For $s\simgt0.8$ these corrections cannot be calculated 
reliably \cite{BUIS} ($s_m=0.65$ of that paper corresponds to $s\approx0.8$) 
which manifests itself in the negative values of the expression 
(\ref{eqn:diffwidth}) for $s\rightarrow1$. To estimate the uncertainty 
introduced by this factor we have computed 
$BR(\bar B\rightarrow X_sl^+l^-)_{0.65<s<s_{\rm max}}$ switching off the 
$1/m_b^2$ corrections in (\ref{eqn:diffwidth}) for $s>0.8$. At $\mu_b=2.5$ 
GeV this gives 
$BR(\bar B\rightarrow X_sl^+l^-)_{0.65<s<s_{\rm max}}=2.66\times10^{-7}$.
The difference of order 15\% between this result and (\ref{eqn:BRllhigh}) 
can be interpreted as the uncertainty associated with the $1/m_b^2$ 
corrections. Adding all uncertainties in quadratures we finally assign to 
the result (\ref{eqn:BRllhigh}) the uncertainty of order $\pm20$\%. 

\section{Scalar flavour changing neutral currents}

Even in the MFV MSSM with $\tan\beta\gg1$ ordinary one loop corrections 
involving charginos and stops can generate substantial FV couplings of 
neutral Higgs bosons to the down-type quarks ($q=s,d$) 
\cite{CHPO1,HAPOTO,BAKO,CHSL}.
For sparticles sufficiently heavier than the charged Higgs boson (which 
sets the mass scale of the MSSM Higgs sector, as in the MSSM for 
$M_{H^+}\simgt200$ GeV $M_H\approx M_A\approx M_{H^+}$) the effects of 
these FV couplings can be described by the local Lagrangian of the form:
\begin{eqnarray}
{\cal L}^{\rm eff}=-\bar q_L\left[X_{\rm LR}\right]^{qb} b_R(H^0-iA^0)
-\bar q_R\left[X_{\rm RL}\right]^{qb} b_L(H^0+iA^0) + {\rm H.c.}~,
\label{eqn:FVcouplings}
\end{eqnarray}
where in the so-called approximation of unbroken $SU(2)\times U(1)$ 
symmetry the amplitudes $\left[X_{\rm LR}\right]^{qb}$ are given by
\cite{BUCHROSL3}
\begin{eqnarray}
\left[X_{\rm LR}\right]^{qb}\approx-{g^3_2\over4}{m_b\over M_W}
\left({m_t\over M_W}\right)^2 {\tan^2\beta ~V^\ast_{tq}V_{tb}
\over(1+\tilde\epsilon_b\tan\beta)
(1+\epsilon_0\tan\beta)}~\epsilon_Y~.
\label{eqn:XLR}
\end{eqnarray}
The factors $\epsilon_Y\sim{\cal O}(1/16\pi^2)$, $\epsilon_0$ and
$\tilde\epsilon_b$ (see ref. \cite{BUCHROSL3} for the  analytical 
expressions) depend on sparticle mass parameters; in particular,
$\epsilon_Y$ is directly proportional to the mixing of left and right 
stops, that is to the parameter $A_t$ \cite{CHSL}. The factors 
$\epsilon_0$ and $\tilde\epsilon_b$ which depend on both, $\alpha_s$ 
and the top Yukawa couplings, ensure proper resummation of the 
$(\tan\beta)$-enhanced terms from all orders of the perturbation
expansion \cite{HAPOTO,BAKO,CHSL,ISRE1,BUCHROSL3,DEPI}. Their signs and 
magnitudes depend directly on the signs of the supersymmetric $\mu$ 
and $A_t$ parameters. Generally, the resummation factors suppress 
the FV couplings for $\mu>0$ \cite{ISRE1} and enhance them for $\mu<0$
\cite{BUCHROSL3}. The amplitudes $\left[X_{\rm RL}\right]^{qb}$ of the 
transitions $b_L\rightarrow s_R(d_R)$ are given by similar expressions 
but with $m_b$ replaced by $m_{s(d)}$ and are, therefore, suppressed 
(but are, nevertheless, important for the $B^0_s$-$\bar B^0_s$ mixing
\cite{BUCHROSL1,BUCHROSL2,CHRO}). The approximate formula (\ref{eqn:XLR}) 
captures the main qualitative features of the FV couplings generated in
the MFV MSSM. For more accurate estimates of their magnitude and 
dependences on the MSSM parameters one has to use, however, more 
complicated approach developed in ref. \cite{BUCHROSL3} which combines 
the resummation of the $(\tan\beta)$-enhanced terms with the complete 
diagramatic 1-loop calculation. In principle, for $M_{\rm SUSY}\gg M_W$ one 
should also take into account that the couplings (\ref{eqn:FVcouplings}) 
are generated in the process of integrating out heavy sparticles at some 
scale $\mu_S\sim M_{\rm SUSY}$ and should be evolved down to the matching 
scale $\mu_0$ using the RGEs similar to the RGEs for the quark Yukawa 
couplings in the SM
\begin{eqnarray}
\mu {d\over d\mu}\left[X_{\rm LR}\right]^{qb}=
-8{\alpha_s\over4\pi}\left[X_{\rm LR}\right]^{qb}+\dots
\end{eqnarray}
where we have retained only the effects of the QCD renormalization. As a 
result, the couplings $\left[X_{\rm LR}\right]^{qb}$ would be multiplied 
by the factor $[\alpha_s(\mu_0)/\alpha_s(\mu_S)]^{4/7}$, equal (for 
$\mu_0=m_t$) 1.073 for $\mu_S=500$ and 1.12 for $\mu_S=1000$ GeV. To take
consistently such effects into account one would have also to determine
sparticle couplings at the scale $\mu_S$ (and use them to compute the
amplitudes $\left[X_{\rm LR}\right]^{qb}$). Since for the correlations
discussed in section 4 only the values of $\left[X_{\rm LR}\right]^{qb}$
at $\mu_0$ matter we will simply assume that sparticles are integrated
out at the same scale $\mu_0=m_t$.

With the FV couplings (\ref{eqn:XLR}) the tree-level exchanges of $H^0$ 
and $A^0$ generate at the scale $\mu_0$ Wilson coefficients of the 
${\cal O}_S^l$ and ${\cal O}_P^l$ operators
\begin{eqnarray}
C^{l(1)}_S(\mu_0)=-{g^4_2\over8M_A^2}
{m_l m_b^{\overline{\rm MS}}(\mu_0)\over M_W^2}
\left({m_t\over M_W}\right)^3 {\tan^3\beta ~V^\ast_{tq}V_{tb}
\over(1+\tilde\epsilon_b\tan\beta)
(1+\epsilon_0\tan\beta)}\epsilon_Y\approx-C_P^{l(1)}(\mu_0)~.
\label{eqn:CSCP}
\end{eqnarray}
Note that the expressions for $C_S^{l(1)}$ and $C_P^{l(1)}$ through 
their dependence (via $\epsilon_0$ and $\tilde\epsilon_b$) on the 
coupling constants $\alpha_s$ and $\alpha_t\equiv y_t^2/4\pi$ (where 
$y_t$ is the top-quark Yukawa coupling) resumm terms of order 
$\alpha_s^n\alpha_t^m\tan^{n+m}\beta$ ($n,m\geq0$) from all orders
of perturbation theory. Since the operators $m_b{\cal O}_{S,P}$, are 
renormalization scale invariant with respect to the strong interactions, 
the QCD evolution of 
\begin{eqnarray}
\tilde C_{S,P}^l\equiv 
C^{l(1)}_{S,P}+{\alpha_s\over4\pi}C^{l(2)}_{S,P}+\cdots
\end{eqnarray}
reduces to the multiplication of $\tilde C_{S,P}^l(\mu_0)$ by the factor
$[m_b(\mu_b)^{\overline{\rm MS}}/m_b^{\overline{\rm MS}}(\mu_0)]$.  
If as in (\ref{eqn:CSCP}) 
$C^{l(1)}_{S,P}(\mu_0)\propto m_b^{\overline{\rm MS}}(\mu_0)$
the dependence on  $m_b^{\overline{\rm MS}}$
of the formula for $BR(B^0_q\rightarrow l^+l^-)$ cancels against 
the factor $1/m_b^{\overline{\rm MS}}(\mu_b)$ 
present in the matrix element of the ${\cal O}_{S,P}$ operators
\cite{BUJAUR}:
\begin{eqnarray}
\langle0|\bar q_Lb_R(\mu_b)|\bar B^0_q\rangle
=i\hat F_{B_q}{M^2_{B^0_q}\over 
m^{\overline{\rm MS}}_b(\mu_b)+m^{\overline{\rm MS}}_q(\mu_b)}
\approx i\hat F_{B_q}{M^2_{B^0_q}\over m^{\overline{\rm MS}}_b(\mu_b)}
\end{eqnarray}
Complete ${\cal O}(\alpha_s)$ calculation of the scalar operators 
contribution to $BR(B^0_q\rightarrow l^+l^-)$ in the MSSM would therefore 
require only computing higher order corrections to the matching conditions 
at the scale $\mu_0$, that is to resumm all contributions to $C_S^{l(2)}$ 
and $C_P^{l(2)}$ of order $\alpha_s(\alpha_s^n\alpha_t^m\tan^{n+m}\beta)$ 
for $n,m\geq0$. 

One can also take a more general point of view and assume that the scalar
operators ${\cal O}_{S,P}^l$ are generated at the scale $\mu_0$ by some
yet unknown physics and investigate their effects on the 
$b\rightarrow sl^+l^-$ and $b\rightarrow dl^+l^-$ transitions without any 
reference to the more fundamental theory, treating the Wilson coefficients 
$\tilde C_{S,P}^l$ as free parameters. Assuming dominance of the scalar 
${\cal O}_{S,P}^l$ operators, the formula for 
$\Gamma(B^0_q\rightarrow l^+l^-)$ \cite{CHSL,BOEWKRUR} takes the form 
\begin{eqnarray}
\Gamma(B^0_q\rightarrow l^+l^-)\approx M_{B^0_q}
{\left(G_F\alpha_{\rm em}M_{B^0_q}\hat F_{B_q}\right)^2\over64\pi^3}
\left({M_{B^0_q}\over m^{\overline{\rm MS}}_b(\mu_b)}\right)^2
|V_{tq}^\ast V_{tb}|^2
\left\{\left|\tilde C^l_S(\mu_b)\right|^2+
\left|\tilde C^l_P(\mu_b)\right|^2\right\}\nonumber
\end{eqnarray}
that is
\begin{eqnarray}
BR(B^0_s\rightarrow l^+l^-)\approx4.27\times10^{-7}
\left({\hat F_{B_s}\over238 ~{\rm MeV}}\right)^2
\left|{V_{ts}^\ast V_{tb}\over0.04}\right|^2
\left({4.2 ~{\rm GeV}\over m^{\overline{\rm MS}}_b(\mu_b)}\right)^2
\nonumber\\
\times{1\over2}\left\{\left|\tilde C^l_S(\mu_b)\right|^2+
\left|\tilde C^l_P(\mu_b)\right|^2\right\}\phantom{aaaaaaaaaaaaa}
\end{eqnarray}
The recent CDF upper limit \cite{NAKAO} 
$BR(B^0_s\rightarrow\mu^+\mu^-)<0.95\times10^{-6}$ at 90 \% C.L. 
sets therefore the stringent bound \cite{BOEWKRUR}
\begin{eqnarray}
{1\over2}\left\{
\left|\tilde C^\mu_S(\mu_b)\right|^2 + 
\left|\tilde C^\mu_P(\mu_b)\right|^2\right\}
\simlt2.2\times\left({238 ~{\rm MeV}\over\hat F_{B_s}}\right)^2
\left({m^{\overline{\rm MS}}_b(\mu_b)\over4.2 ~{\rm GeV}}\right)^2
\label{eqn:Cs_lim}
\end{eqnarray}
Similar bound can be also derived for $\tilde C^e_{S,P}(\mu_b)$ by using 
the corresponding experimental upper limit 
$BR(B^0_s\rightarrow e^+e^-)<5.4\times10^{-5}$ \cite{PDG} but it is two 
orders of magnitude weaker. Analogous bounds on the (universal under the 
assumption of MFV) Wilson coefficients $\tilde C^{e,\mu}_{S,P}(\mu_b)$ 
that can be derived from the experimental upper limits  
$BR(B^0_d\rightarrow\mu^+\mu^-)<1.6\times10^{-7}$ \cite{BABAR_Bdmm,NAKAO} 
and $BR(B^0_d\rightarrow e^+e^-)<8.3\times10^{-7}$ \cite{PDG} are less 
interesting as they depend on the value of $|V_{td}|$, determination of 
which can be also affected by the new physics that gives rise to the scalar 
operators \cite{CHRO}. 

As follows from the formula (\ref{eqn:CSCP}), in the MSSM 
$\tilde C^l_{S,P}\approx C^{l(1)}_{S,P}\propto m_l$, so that the effects of 
the scalar operators can be measurable only for the $\mu^+\mu^-$ and 
$\tau^+\tau^-$ channels (the latter being very difficult for experimental 
searches so that at present no limit on $BR(B^0_s\rightarrow\tau^+\tau^-)$ 
is available). For $\tan\beta\sim40-50$, substantial stop mixing and $\mu<0$, 
when the resummation of the leading $\tan^n\beta$ terms enhances the FV 
violating couplings, $|C^{\mu(1)}_S|\approx|C^{\mu(1)}_P|$ could be as 
large as $\sim10$ leading to $BR(B^0_s\rightarrow\mu^+\mu^-)\sim10^{-5}$ 
\cite{CHSL,BUCHROSL3}. The bound (\ref{eqn:Cs_lim}) eliminates 
therefore a large portion of the general MSSM parameter space.
Moreover, as has been demonstrated in \cite{BUCHROSL2,BUCHROSL3}, in such
cases also the contribution of the FV couplings of the neutral MSSM Higgs 
bosons to the $B^0_s$-$\bar B^0_s$ mass difference $\Delta M_s$ is large and 
the experimental limit $(\Delta M_s)^{\rm exp}\simgt14/$ps \cite{OSC_BsBs} 
becomes in most cases more constraining (see the next section). 

It should be stressed, however, that the the bounds like (\ref{eqn:Cs_lim}) 
are completely independent of the specific way of generation of the 
coefficients $|\tilde C^l_{S,P}|$ and are valid generally, and not only 
in supersymmetry.\footnote{The bound (\ref{eqn:Cs_lim}) is valid also if
the new physics, which gives rise to nonzero $\tilde C^l_{S,P}$ involves 
sources of FV other than the CKM matrix, provided the coefficients 
$\tilde C^l_{S,P}$ are
(superficially) normalized as in (\ref{eqn:Heff}). More generally, for a 
given lepton pair $l^+l^-$ the experimental upper limits on 
$BR(B^0_s\rightarrow l^+l^-)$ and $BR(B^0_d\rightarrow l^+l^-)$ set then 
independent bounds on the products (assuming that the Wilson coefficients
are still normalized as in (\ref{eqn:Heff}))
$|V_{tq}^\ast V_{tb}|^2\left\{\left|\tilde C^l_S\right|^2+
\left|\tilde C^l_P\right|^2\right\}$ for $q=s$ and $q=d$, respectively, 
which can be directly used to constrain the maximal possible effects of the 
scalar operators in inclusive or exclusive $\bar B\rightarrow X_sl^+l^-$ 
and $\bar B\rightarrow X_dl^+l^-$ decays.}
In particular, one can imagine that the operators ${\cal O}_{S,P}^l$ are 
not due to to the neutral Higgs boson exchanges between the FV violating 
down-type quark vertices and the leptonic vertices in which case sizeable 
effects of the scalar operators ${\cal O}_{S,P}^l$ could be present in any 
of the $b\rightarrow s(d)l^+l^-$ transitions (for any lepton) and not 
accompanied by large contributions to the $B^0_s$-$\bar B^0_s$ mixing 
amplitude as in the MSSM. For this reason, the remaining analysis of this 
section will be done in a general framework. We will return to the MSSM 
only in the next section.

The general bound (\ref{eqn:Cs_lim}) on $|\tilde C^\mu_{S,P}|$ allows for an 
immediate estimate of the impact, the scalar operators ${\cal O}^\mu_{S,P}$ 
may have on the rate of the inclusive process $B\rightarrow X_s\mu^+\mu^-$.
Similar estimates can be also made for  $B\rightarrow X_se^+e^-$ and
$B\rightarrow X_s\tau^+\tau^-$ processes. From the formula 
(\ref{eqn:diffwidth}) for the contribution of the scalar operators to the 
differential rate we get \cite{XIYA,WAAT}:
\begin{eqnarray}
{d\over ds}\Delta BR(\bar B\rightarrow X_s\mu^+\mu^-)
\approx{BR(\bar B\rightarrow X_ce\nu_e)\over
\left(1-{2\alpha_s(m_b)\over3\pi}h(0)\right)}{1\over C}
\left|{V_{ts}^\ast V_{tb}\over V_{cb}}\right|^2
\left({\alpha_{\rm em}\over2\pi}\right)^2\phantom{aaaaaa}\nonumber\\
\times(1-s)^2\left\{{3\over2}s\left|\tilde C^\mu_S(\mu_b)\right|^2
+{3\over2}s\left|\tilde C^\mu_P(\mu_b)\right|^2
+6{m_\mu\over m_b}\tilde C^\mu_P(\mu_b)C^{\rm eff}_{10}(s,\mu_b)\right\}
\label{eqn:delta_diffwidth}
\end{eqnarray}
where we have used the normalization to the width of the semileptonic
charmless decays and for simplicity dropped the nonperturbative correction 
factor appearing in the denominators of the formula (\ref{eqn:Ctrick}). As 
remarked below the formulae (\ref{eqn:tildeC}), the contribution of the 
operators ${\cal O}_{S,P}^l$ to the inclusive rate 
$BR(\bar B\rightarrow X_s\mu^+\mu^-)$ depends on the choice of the 
renormalization scale $\mu_b$. Since following ref.~\cite{GAMI} we use 
$m_b^{1S}=4.69$ GeV leading to
$m_b^{\overline{\rm MS}}(m_b^{\overline{\rm MS}})\approx4.2$ GeV for the 
value of the running $b$-quark mass, in what follows we will treat as free
parameters the Wilson coefficients $\tilde C^\mu_{S,P}$ taken at $\mu_b=4.2$ 
GeV. The uncertainty related to the variation of the scale 
$\mu_b\rightarrow\mu_b^\prime$ in the formula (\ref{eqn:delta_diffwidth})
is then roughly (ascribing for the estimation purpose to the interference 
term the same $\mu_b$ dependence as have the other two terms) given by 
$[m_b^{\overline{\rm MS}}(\mu_b^\prime)/m_b^{\overline{\rm MS}}(\mu_b)]^2$ 
and is 
estimated to be $^{+22}_{-25}$\%. This uncertainty has to be, of course,
combined with the ones stemming from unknown electromagnetic corrections
and the $C$-factor (\ref{eqn:Cfact}).
Inserting numbers in the formula (\ref{eqn:delta_diffwidth}) we get
\begin{eqnarray}
{d\over ds}\Delta BR(\bar B\rightarrow X_s\mu^+\mu^-)
\approx4.7\times10^{-7}\phantom{aaaaaaaaaaaaaaaaaaaaaaaaaa}\nonumber\\
\times(1-s)^2\left\{s\left|\tilde C^\mu_S(\mu_b)\right|^2
+s\left|\tilde C^\mu_P(\mu_b)\right|^2
+4{m_\mu\over m_b}\tilde C^\mu_P(\mu_b)C^{(1)}_{10}\right\}~.
\end{eqnarray}

Integrating over the full $(0,1)$ range of $s$ and taking into account the 
limit (\ref{eqn:Cs_lim}) with $m_b^{\overline{\rm MS}}(\mu_b)=4.2$ GeV for 
$\mu_b=4.2$ GeV we obtain the estimate of the maximal possible contribution 
of the scalar operators to the ``non-resonant'' branching ratio:
\begin{eqnarray}
\Delta BR(\bar B\rightarrow X_s\mu^+\mu^-)_{\rm nonres}
\simlt1.7\times f\times\left(1\pm0.5\times\sqrt{r/f}\right)\times10^{-7}
\label{eqn:deltaBRfull}
\end{eqnarray}
where
\begin{eqnarray}
f\equiv\left({238 ~{\rm MeV}\over\hat F_{B_s}}\right)^2~,
\phantom{aaaa}0.78<f<1.32\phantom{aaaaaaaa}
\end{eqnarray}
and the factor
\begin{eqnarray}
0\leq r\leq 2
\end{eqnarray}
depends on the relative magnitudes of $|\tilde C^\mu_P|$ and 
$|\tilde C^\mu_S|$: $r=0$ for $|\tilde C^\mu_P|=0$ and $r=2$ for 
$|\tilde C^\mu_S|=0$; for $|\tilde C^\mu_P|=|\tilde C^\mu_S|$, as in the 
MSSM, $r=1$. The $\pm$ refers to the two possible signs of the interference 
term depending on the sign of $\tilde C_P^\mu$ (the interference is 
constructive for $\tilde C_P^\mu<0$). We have used the approximate SM value 
$\tilde C^{\rm eff}_{10}(s,\mu_b)\approx C_{10}^{(1)}\approx-4.2$ and 
$m_b=m_b^{\rm pole}=4.8$ GeV in the interference term. Thus, the maximal 
effect of the scalar operator is $3.7\times10^{-7}$ for $f=1.32$ and $r=2$ 
($2.55\times10^{-7}$ for $f=r=1$). Comparing with the SM result 
(\ref{eqn:BRmmnonres}) we conclude that the maximal contribution of the 
scalar operators allowed by the CDF limit (\ref{eqn:mainlimit}) is at 
most at the level of 8\%  
for $f=1.32$, $r=2$ (5\% for $f=r=1$), that is, substantially smaller the 
estimated uncertainty of the SM prediction. This is in sharp contrast with 
the findings of ref. \cite{WAAT}, where it has been claimed that even 
within the so-called minimal SUGRA framework the ratio 
$BR(B^0_s\rightarrow\mu^+\mu^-)/BR(B^0_s\rightarrow e^+e^-)$
can reach values as big as 2-3, corresponding to the contribution
of the scalar operators as large as 100-200\%.

For the branching ratio integrated over the range $0.05<s<0.25$ we find 
\begin{eqnarray}
\Delta BR(\bar B\rightarrow X_s\mu^+\mu^-)_{0.05<s<0.25}\simlt
0.43\times f\times\left(1\pm0.88\times\sqrt{r/f}\right)\times10^{-7}
\label{eqn:deltaBRlow}
\end{eqnarray}
that is, the maximal effect is again of order 8\% for $f=1.32$, $r=2$ (5.5\% 
for $f=r=1$), much smaller than the estimated uncertainty of the SM prediction 
for this range. For the range of $s$ above the $\bar cc$ resonances the limit 
(\ref{eqn:mainlimit}) implies:
\begin{eqnarray}
\Delta BR(B\rightarrow X_s\mu^+\mu^-)_{0.65<s<1}\simlt
0.22\times f\times\left(1\pm0.16\times\sqrt{r/f}\right)\times10^{-7}
\label{eqn:deltaBRhigh}
\end{eqnarray}
For this $s$ range the maximal possible contribution of the scalar 
operators increases the branching fraction by $\sim15$\% for $f=1.32$, 
$r=2$ (11\% for $f=r=1$), that is again the effects of the scalar 
operators are not greater than the estimated uncertainty of the SM 
prediction.\footnote{With the old limit 
$BT(B^0_s\rightarrow\mu^+\mu^-)<2\times10^{-6}$ \cite{CDF_Bsmm} the effects 
of the scalar operators in this range of $s$ could be almost twice as big as 
the estimated uncertainty.}
Estimates of $\Delta BR(B\rightarrow X_se^+e^-)$ 
can be also obtained in a similar manner. 

Experimentally first measured were the exclusive $B$ decay modes
$\bar B\rightarrow Kl^+l^-$ and  $\bar B\rightarrow K^\ast l^+l^-$
\cite{BKLLexp}. For $\bar B\rightarrow Kl^+l^-$, which will be of interest 
for us here,\footnote{As analyzed in ref. \cite{BOEWKRUR}, the contribution 
of the scalar operators to $BR(\bar B\rightarrow K^\ast\mu^+\mu^-)$ 
is too small to be interesting.}
the recent results for the ``nonresonant'' rates are \cite{NAKAO}: 
$BR(\bar B\rightarrow K\mu^+\mu^-)
=(4.8^{+1.5}_{-1.3}\pm0.3\pm0.1)\times10^{-7}$ and
$BR(\bar B\rightarrow Kl^+l^-)=(4.8^{+1.0}_{-0.9}\pm0.3\pm0.1)\times10^{-7}$
averaged over $e$ and $\mu$ (BELLE) and 
$BR(\bar B\rightarrow K\mu^+\mu^-)
=(4.8^{+2.5}_{-2.0}\pm0.4)\times10^{-7}$ and
$BR(\bar B\rightarrow Kl^+l^-)=(6.9^{+1.5}_{-1.3}\pm0.6)\times10^{-7}$
(BaBar).
The main uncertainty of the theoretical $BR(\bar B\rightarrow Kl^+l^-)$ 
calculation is related to the determination of the nonperturbative matrix 
elements of the relevant operators between the initial and final meson states.
Different techniques used for this purpose resulted in the SM predictions
for this branching fraction spanning the range $(3.0-6.9)\times10^{-7}$
\cite{COFASASC,ALBAHAHI,ALLUGRHI}. Within the experimental errors the new 
experimental results are in fair agreement with the SM-based NNLO theoretical 
estimate given by Ali et al. \cite{ALLUGRHI}:
$BR(\bar B\rightarrow Kl^+l^-)_{\rm nonres}=(3.5\pm1.2)\times10^{-7}$.
Substantial lowering of the SM prediction compared to the earlier one
of Ali et al. (based on the NLO calculation) \cite{ALBAHAHI}, 
$BR(\bar B\rightarrow Kl^+l^-)_{\rm nonres}=(5.7\pm1.2)\times10^{-7}$
was mainly due to the superficial lowering of  
values of the formfactors parametrizing the operator matrix elements. 
This was motivated by the fact that the $q^2=0$ value of the 
$T_1(q^2)$ formfactor obtained using the so-called QCD light cone sum 
rules (LCSR) gave, compared to the data, too high a branching fraction 
for the $\bar B\rightarrow K^\ast\gamma$ mode \cite{ALPA}, suggesting 
that the LCSR method systematically overestimates the formfactors.

The contribution of the scalar operators to the branching fraction
$BR(\bar B\rightarrow K\mu^+\mu^-)_{\rm nonres}$ has been analyzed 
in ref.~\cite{BOEWKRUR}. At that time only the upper limit 
$BR(B^+\rightarrow K^+\mu^+\mu^-)<5.2\times10^{-6}$ was available
\cite{AFFexp}, so the conclusion of ref. \cite{BOEWKRUR} was that the 
constraint imposed on $|\tilde C_S^\mu|^2+|\tilde C_P^\mu|^2$ by the limit
$BR(B^0_s\rightarrow \mu^+\mu^-)<2.6\times10^{-6}$ was significantly 
stronger than the one that could be obtained from the limit on
$BR(\bar B\rightarrow K^+\mu^+\mu^-)$. With the new numbers the situation 
is somewhat different and we summarize it below. 

The scalar operators contribution to the nonresonant branching ratio
can be written as \cite{BOEWKRUR}
\begin{eqnarray}
{d\over dq^2}\Delta Br(\bar B\rightarrow Kl^+l^-)_{\rm nonres}
={\tau_B\over\pi}\left({G_F\alpha_{\rm em}\over16\pi^2}\right)^2
{|V_{ts}^\ast V_{tb}|^2\over M^3_B}\lambda^{1/2}(q^2,M^2_B,M^2_K)
\beta_l(q^2)\nonumber\\
\times\left\{q^2\beta_l^2(q^2)|\delta F_S|^2+q^2|\delta F_P|^2 
+ 2q^2{\rm Re}(F_P^\ast\delta F_P)\phantom{aa}\right.
\label{eqn:dBRKll}\\
\left.+ 2m_l(M_B^2-M^2_K+q^2){\rm Re}(F_A^\ast\delta F_P)\right\}
\phantom{aaaa}\nonumber
\end{eqnarray}
where $q^2$ is the physical lepton pair invariant mass, 
$\beta_l(q^2)=\sqrt{1-4m^2_l/q^2}$ and 
\begin{eqnarray}
&&\delta F_{S,P}={1\over2}{C_{S,P}^l(\mu_b)\over 
m_b^{\overline{\rm MS}}(\mu_b)}
(M_B^2-M^2_K)~f_0(q^2)\nonumber\\
&&F_A = C_{10}^{\rm eff}f_+(q^2)\\
&&F_P = m_l C_{10}^{\rm eff}\left\{
{M^2_B-M^2_K\over q^2}\left[f_+(q^2)-f_0(q^2)\right]-f_+(q^2)\right\}
\nonumber
\end{eqnarray}
The coefficient $C_{10}^{\rm eff}$ differs from 
$\tilde C_{10}^{\rm eff}(s,\mu_b)$ given in eq. (\ref{eqn:tildeC}) by 
setting to zero
the functions $\omega_9(s)$ (the effects of $\omega_9(s)$ are supposed 
to be taken into account in the formfactors $f_0(q^2)$ and $f_+(q^2)$).
Note that $C_{10}^{\rm eff}$ \cite{BOMIUR}, and hence the whole formula 
(\ref{eqn:dBRKll}), is independent of the renormalization scale $\mu_b$.
Following the recipe of ref. \cite{ALLUGRHI} for the central values of the 
formfactors $f_0(q^2)$ and $f_+(q^2)$, as well as for $f_T(q^2)$ appearing 
below, in eq. (\ref{eqn:fv}), we use their lowest values obtained
within the LCSR approach which amounts to using the formula (3.7) of 
\cite{ALBAHAHI} with the parameters collected in Table V of that paper.
At the same time, again following ref. \cite{ALLUGRHI}, we ascribe to the 
values of the formfactors the uncertainty of order $15$\%. The formfactors 
introduce therefore in the results for 
$(d/dq^2)\Delta Br(\bar B\rightarrow Kl^+l^-)_{\rm nonres}$
the largest (barring the discussion how big 
errors are introduced by using the effective Lagrangian with non-local 
coefficients $C^{\rm eff}_9(q^2)$, $C^{\rm eff}_7(q^2)$, for the exclusive 
process) uncertainty of order 30\%. 

Integrating over $q^2$ in the kinematical limits $4m^2_\mu<q^2<(M_B-M_K)^2$ 
and assuming that the new physics contribution to Wilson coefficients other 
than $\tilde C^\mu_{S,P}$ is negligible we obtain for the dimuon mode
\begin{eqnarray}
\Delta Br(\bar B\rightarrow K\mu^+\mu^-)_{\rm nonres}
\approx6.36\times10^{-8}\times
\left\{a\left(\left|\tilde C^\mu_S\right|^2+
\left|\tilde C^\mu_P\right|^2\right)
-b~\tilde C^\mu_P\right\}
\end{eqnarray}
where for $\mu_b=4.2$ GeV $a=0.30\pm0.10$ and $b=0.19\pm0.06$. The 
uncertainties of $a$ and $b$ are due to the uncertainties of the 
formfactors $f_0(q^2)$ and $f_+(q^2)$. Through the formfactors the total 
uncertainty of the scalar operators contribution is obviously strongly 
correlated with the uncertainty of the SM prediction. Sticking to the 
central values of $a$ and $b$ and taking maximal values of 
$|\tilde C_S^\mu|$ and $|\tilde C_P^\mu|$ allowed by the bound 
(\ref{eqn:Cs_lim}) we get
\begin{eqnarray}
\Delta Br(\bar B\rightarrow K\mu^+\mu^-)_{\rm nonres}
\simlt 0.8\times f\times
\left(1\pm0.45\sqrt{r/f}\right)\times10^{-7}\label{eqn:deltaBKmumu}
\end{eqnarray}
that is, the maximal possible contribution of the scalar operators to the 
nonresonant branching fraction can be (for $\tilde C^\mu_P<0$, 
$\tilde C^\mu_S=0$, and the lowest possible value of $\hat F_{B_s}$, i.e. 
for the $+$ sign, $r=2$, $f=1.32$) as large as $1.7\times10^{-7}$, roughly 
of the same magnitude as the error of the experimental result and
$1.5$ times bigger than the estimated \cite{ALBAHAHI,BOEWKRUR,ALLUGRHI}) 
uncertainty ($\sim1.2\times10^{-7}$) of the SM prediction. Similar estimates 
can be also done for $\Delta Br(\bar B\rightarrow Ke^+e^-)_{\rm nonres}$.

Finally, an experimentally interesting quantity \cite{BOEWKRUR,CHCOGAJO} 
may be the integrated over $q^2$ forward-backward lepton asymmetry 
measured in this decay given by\footnote{Of interest can be also unintegrated
differential asymmetry \cite{DEOLVO}.}
\begin{eqnarray}
A_{\rm FB}={\tau_B\over BR(\bar B\rightarrow K\mu^+\mu^-)}
\left({G_F\alpha_{\rm em}\over16\pi^2}\right)^2
{|V_{ts}^\ast V_{tb}|^2\over\pi M^3_B}\phantom{aaaaaa}\nonumber\\
\times\int dq^2 ~m_l\lambda(q^2,M^2_B,M^2_K)~
\beta_l^2(q^2)~{\rm Re}(F_V^\ast\delta F_S)
\end{eqnarray}
where 
\begin{eqnarray}
F_V= C_9^{\rm eff}(\mu_b)f_+(q^2) +2m_b C_7^{\rm eff}(\mu_b)
{f_T(q^2)\over M_B+M_K}\label{eqn:fv}
\end{eqnarray}
with $C_9^{\rm eff}$ and $C_7^{\rm eff}$ differing from 
$\tilde C_9^{\rm eff}$ and $\tilde C_7^{\rm eff}$ of eqs. 
(\ref{eqn:tildeC}) by setting to zero\footnote{for $q^2/m_b^2>0.25$ 
we also set to zero the functions $F_i^{(7,9)}$.}
the functions $\omega_9(s)$ and $\omega_7(s)$. The asymmetry $A_{\rm FB}$ 
vanishes in the SM in which $F_S=\delta F_S=0$. For the dimuon channel,
integrating over the whole $q^2$ range and using $\mu_b=4.2$ GeV we get
\begin{eqnarray}
A_{\rm FB}\approx{1\over BR(\bar B\rightarrow K\mu^+\mu^-)} \times
(4.9\pm1.6)\times\tilde C_S^\mu\times10^{-9} \nonumber\\
\simlt\pm
\left[{4.8\times10^{-7}\over BR(\bar B\rightarrow K\mu^+\mu^-)}\right] 
\times(1.5\pm0.5)\times\sqrt{r^\prime f} ~\%\label{eqn:AFB}
\end{eqnarray}
where $0<r^\prime<2$ ($r^\prime=0$ for $C_S^\mu=0$ and $r^\prime=2$ for 
$C_P^\mu=0$; for $|\tilde C_S^\mu|=|\tilde C_P^\mu|$, as in the MSSM with, 
$r^\prime=1$). The uncertainty of this result being dominated by the 30\% 
uncertainty arising from the formfactors $f_+(q^2)$ and $f_T(q^2)$), is of 
course strongly correlated with the uncertainty of the total branching 
ratio. Still, the maximal possible asymmetry allowed by the limit 
(\ref{eqn:Cs_lim}) is of the order of a percent and may be detectable in 
the future.

We conclude that given the experimental limit (\ref{eqn:mainlimit}), the 
effects of the scalar operators in the inclusive process are typically of 
order $5-15$\%, always smaller than the estimated uncertainty of the SM NNLO 
prediction. On the other hand, the maximal allowed contribution of the scalar 
operators to $BR(\bar B\rightarrow K\mu^+\mu^-)$, although larger than the 
estimates of the theoretical uncertainty of the SM prediction made in 
\cite{ALBAHAHI,BOEWKRUR,ALLUGRHI}, is only roughly of the order of the present 
experimental error. While the latter can shrink in the near future, the 
spread of the different SM based theoretical 
predictions and the problems with the formfactor values obtained using the 
QCD LCSR may suggest that the true uncertainty of the SM prediction is
larger than estimated in \cite{ALBAHAHI,BOEWKRUR,ALLUGRHI}, thus preventing 
the reliable comparison of the theoretical predictions with the data.
The forward backward asymmetry of the muon distribution, if detected
in the high statistic data, could be also indicative of the scalar
operators contribution (the asymmetry vanishes if only the SM operators
contribute) but its translation into the values of $\tilde C_S^\mu$ and
$\tilde C_P^\mu$ depends on the formfactors too. Thus, before the status of
the formfactors is clarified and the errors associated with them
reliably estimated the exclusive mode $\bar B\rightarrow K\mu^+\mu^-$,
although potentially interesting, will not be able to put constraints on 
the coefficients $\tilde C_S^\mu$ and $\tilde C_P^\mu$. 

The coefficients $\tilde C_S^\mu$ and $\tilde C_P^\mu$ of the most 
interesting (largest allowed) magnitude cannot be however, as in 
supersymmetry, due to the tree 
level exchanges of the neutral Higgs bosons between the effective quark FV 
vertices and the Higgs-lepton-lepton vertices. As we shall see on the MSSM 
example in the next section, in such a case possible effects of the scalar 
operators (apart from being slightly reduced by the relation 
$|C_S^{l(1)}|\approx|C_P^{l(1)}|$ so that $r=r^\prime=1$) can be further 
constrained by the $\bar B^0_s$-$B^0_s$ mixing.

\section{Correlation with $BR(\bar B\rightarrow X_s\gamma)$ and the 
$\bar B^0_s$-$B^0_s$ mass difference}

In assessing potential effects of the scalar operators in the preceding
section we have ignored the fact that the new physics, which gives rise to 
them, can also modify the remaining Wilson coefficients. In the MFV MSSM
charginos and stops as well as the charged Higgs boson $H^+$ contribute
to $C_{10}(\mu_0)$ and $C_9(\mu_0)$ through the box, $Z^0$-penguin and, 
in the case of $C_9(\mu_0)$, also through the photonic penguin diagrams. 
Likewise the coefficients of the $C_7$ and $C_8$ are modified by loops 
containing these particles. It should be also stressed that supersymmetric
contributions to $C^{(2)}_X(\mu_0)$ in eq. (\ref{eqn:CXexp}) necessary for 
complete NNLO calculations are only partly known for $C_7$ and $C_8$ (and 
only for a scenario with light right-handed stop and charginos) 
\cite{CIDEGAGI} and are unknown for the other coefficients in 
(\ref{eqn:Heff}). Out of the relevant 
for the $b\rightarrow sl^+l^-$ transition Wilson coefficients only the 
modulus of $\tilde C_7^{\rm eff}$ (but not its sign) is rather well 
constrained by the measurement of $BR(\bar B\rightarrow X_s\gamma)$. 
The other coefficients can still accommodate substantial new physics 
contributions. 

If $H^+$ is light - a necessary condition for generating in the MSSM
nonegligible Wilson coefficients of the scalar operators - its contribution
to $\tilde C_7^{\rm eff}$ is substantial and has the same sign as 
the SM contribution. Therefore it must be cancelled out by the chargino-stop
contribution. For $\tan\beta\gg1$ the latter is proportional to $\tan\beta$ 
and can be very large if these particles are light. Its sign depends on the 
sign of $A_t\mu$ and for $A_t\mu>0$ (in our phase convention) it is opposite 
to the sign of the $W^-t$ and $H^+t$ loops so that the cancellation is indeed
possible. Since for $A_t\mu>0$ the Wilson coefficient $C_P^{l(1)}$ is 
negative ($C_S^{l(1)}$ is positive), the requirement that the calculated in 
the MSSM $BR(\bar B\rightarrow X_s\gamma)$ agrees for a light $H^+$ with the 
experimental result necessarily leads to positive contribution of the 
interference term Re$(\tilde C_{10}^{\rm eff} C^{l(1)\ast}_P)$ in the 
formulae (\ref{eqn:diffwidth}) and (\ref{eqn:delta_diffwidth}) (recall 
that the SM contribution to $\tilde C_{10}^{(1)}$ is also negative) 
so that in the estimates (\ref{eqn:deltaBRfull}), (\ref{eqn:deltaBRlow}),
(\ref{eqn:deltaBRhigh}), (\ref{eqn:deltaBKmumu}) and (\ref{eqn:AFB}) the 
$+$ signs apply.

In principle the chargino-stop contribution could even reverse 
the sign of $\tilde C_7^{\rm eff}$ leading to a value of 
$BR(\bar B\rightarrow X_s\gamma)$ compatible with the experimental result. 
The sign of the Re$(\tilde C_7^{\rm eff}\tilde C_9^{{\rm eff}\ast})$ term 
in the formula (\ref{eqn:diffwidth}) would be then changed modifying 
predictions for the $\bar B\rightarrow X_s l^+l^-$ rates. Such situation, 
which could most easily be distinguished by measuring the dilepton 
invariant mass spectrum and the forward backward asymmetry in the 
$BR(\bar B\rightarrow X_s l^+l^-)$ \cite{ALGIMA}, requires light, 
$\sim100$ GeV, charginos 
and stops and, for light $H^+$ and $\tan\beta\gg1$, is strongly fine tuned 
\cite{CHPORb}. Much more natural appears the possibility that charginos and 
stops are rather heavy and their contribution to $\tilde C_7^{\rm eff}$, 
despite substantial stop mixing necessary for generating large 
$C_{S,P}^{l(1)}$, is small, just of the right magnitude (and sign) to cancel 
the contribution of the charged Higgs boson. In such a scenario the value 
of $\tilde C_7^{\rm eff}$ must be close to the one predicted in the SM and 
the contributions of stops and charginos to $C^{(1)}_{10}(\mu_0)$ and 
$C^{(1)}_9(\mu_0)$ is, as we have checked by using the formulae of 
ref. \cite{CHMIWY}, negligible. 

The $H^+$ contribution to $C^{(1)}_{10}(\mu_0)$ and $C^{(1)}_9(\mu_0)$
through the box diagrams, $Z^0$ and photonic penguins has been computed 
in ref. \cite{CHMIWY}. For $\tan\beta\gg1$ these contributions are not 
enhanced and are negligible even for the charged Higgs boson mass as low 
as 200 GeV. As has been found in \cite{LONI,CHSL} the $H^+t$ loops also 
generate the FV couplings (\ref{eqn:FVcouplings}) and the resulting 
contribution to $C_{S,P}^{l(1)}$ grows as $\tan^2\beta$. However, for 
$M_{H^+}\simgt200$ GeV and $\tan\beta\simlt50$ this contribution to the 
coefficients $C_{S,P}^{\mu(1)}$ are roughly two orders of magnitude below 
the upper limit (\ref{eqn:Cs_lim}) and, hence, their impact on the 
$\bar B\rightarrow X_s l^+l^-$ rate can also be neglected.

Thus, for sparticles heavier than, say, 500 GeV, the only sizeable SUSY 
effects in the $b\rightarrow s\mu^+\mu^-$ transitions can  be due
to scalar operators.

As has been observed in \cite{BUCHROSL1,BUCHROSL2,BUCHROSL3}, in the MFV
MSSM whenever the coupling $\left[X_{\rm LR}\right]^{sb}$ (\ref{eqn:XLR}) 
is large, the tree level exchanges of the neutral Higgs bosons $H^0$ and 
$A^0$ between the tree-level effective vertices (\ref{eqn:FVcouplings}) 
give also large negative contribution to the mass difference $\Delta M_s$ 
of the $\bar B^0_s$ and $B^0_s$ mesons. In the so-called approximation of 
unbroken $SU(2)\times U(1)$ symmetry, in which also the formula 
(\ref{eqn:XLR}) is valid, one gets \cite{BUCHROSL3}
\begin{eqnarray}
\delta(\Delta M_s)=-{12.8\over{\rm ps}}
\left[{\tan\beta\over50}\right]^4
\left[{\hat F_{B_s}\over238~{\rm MeV}}\right]^2
\left[{|V_{ts}|\over0.04}\right]^2
\left[{m_b(\mu_0)\over3~{\rm GeV}}\right]
\left[{m_s(\mu_0)\over60~{\rm MeV}}\right]
\nonumber\\
\times\left[{m_t^4\over M_W^2M_A^2}\right]
\left[{16\pi^2\epsilon_Y\over(1+\epsilon_0\tan\beta)
(1+\tilde\epsilon_b\tan\beta)}\right]^2\phantom{aaaa}
\label{eqn:ddms}
\end{eqnarray}
(the analogous contribution to the $\bar B^0_d$-$B^0_d$ mass difference,
being suppressed by the ratio $m_d/m_s$, is negligible).
Typically the couplings 
$\left[X_{\rm LR}\right]^{sb}$ which give rise to $C^{\mu(1)}_{S,P}$
saturating the bound (\ref{eqn:Cs_lim}) lead to 
$\Delta M_s$ below the present lower experimental limit $\sim14/$ps
\cite{OSC_BsBs}. 

\begin{figure}
\epsfig{figure=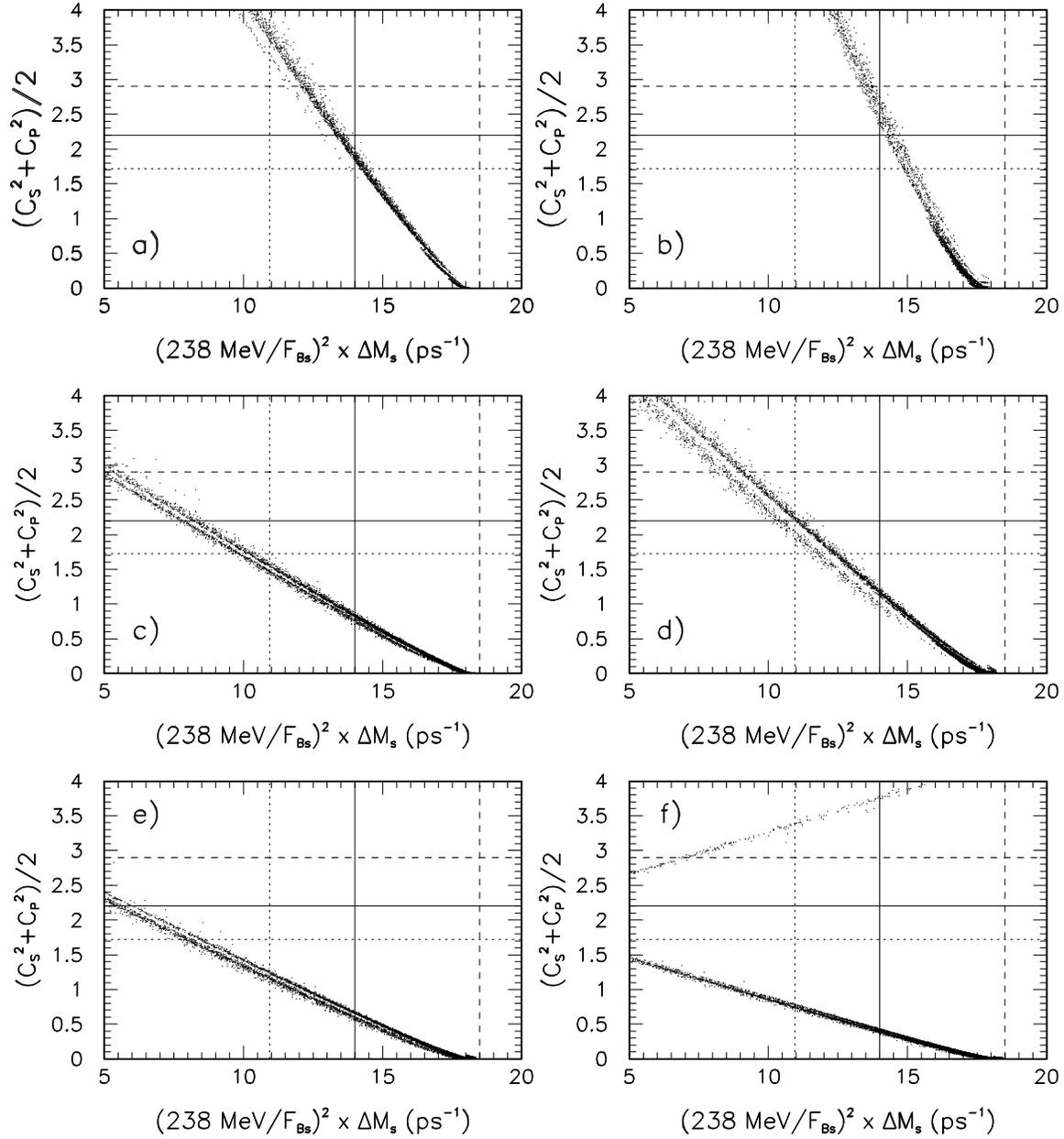,width=\linewidth} 
\vspace{1.0truecm}
\caption{\protect Scatter plots of 
$(1/2)\left(|C^{\mu(1)}_S|^2+|C^{\mu(1)}_P|^2\right)$ versus 
$\Delta M_s$ in the MFV MSSM for sparticle masses greater than $500$ GeV. 
Panels a-f correspond to $(M_A,\tan\beta)$ values (200,40), (200,50), 
(300,40), (300,50), (400,50), (500,50), respectively. Points to the left 
and above the solid lines are for $\hat F_{B_s}=238$ MeV excluded by  
$\Delta M_s>14/$ps and $BR(B^0_s\rightarrow\mu^+\mu^-)<0.95\times10^{-6}$,
respectively. The same constraints for $\hat F_{B_s}=207$ MeV and
269 MeV are shown by dashed and dotted lines, respectively.}
\label{fig:sfcnc1}
\end{figure}

In order to see how the possible effects of the scalar operators 
in the $b\rightarrow s\mu^+\mu^-$ transitions are limited by the
experimental lower bound on the $\bar B^0_s$-$B^0_s$ mass difference
$\Delta M_s$ we present in figs. \ref{fig:sfcnc1}a-\ref{fig:sfcnc1}f 
scatter plots of the combination
$(1/2)\left(|C^{\mu(1)}_S(\mu_b)|^2+|C^{\mu(1)}_P(\mu_b)|^2\right)
\approx|C^{\mu(1)}_S(\mu_b)|^2\approx|C^{\mu(1)}_P(\mu_b)|^2$ for 
$\mu_b=4.2$~GeV versus $\Delta M_s$ calculated using the approach 
developed in \cite{BUCHROSL3} for a few combinations of the parameters 
$(M_A,\tan\beta)$. 
\footnote{In producing these plots we have corrected a bug in our fortran
code which resulted in using in refs. \cite{BUCHROSL2,CHRO,BUCHROSL3}
$C^{\mu(1)}_{S,P}(\mu_0)$ instead of $C^{\mu(1)}_{S,P}(\mu_b)$ in 
calculating $BR(B^0_{s,d}\rightarrow\mu^+\mu^-)$. As a result numerical 
values of this ratios in figures of these references should be rescaled 
upwards roughly by a factor 
$[m^{\overline{\rm MS}}_b(4.2 ~{\rm GeV})/m^{\overline{\rm MS}}_b(m_t)]^2
\approx2.36$.}
The plots have been obtained by scanning over the MFV MSSM parameters 
(in the sense explained in more detail in ref. \cite{BUCHROSL3}) with 
the lower bound on sparticle masses $M_{\rm SUSY}\simgt500$ GeV. More 
specifically, we have scanned the relevant parameters in the ranges 
such that: 500 GeV$<m_{C_1}<1$ TeV, with $0.75<|M_2/\mu|<1.5$, 
$1<m_{\tilde g}/M_2<3$; $0.7<M_{\tilde t_1}/m_{C_1}<1.3$, 
$1.1<M_{\tilde t_2}/M_{\tilde t_1}<1.7$ and $-35^o<\theta_{\tilde t}<35^o$; 
$0.5<M_{\tilde b_R}/m_{\tilde g}<0.9$, with $A_b=A_t$; masses of the first 
two generations have been taken as max$(M_{\tilde b_R}, M_{\tilde t_L})$.
All points, 
for which computed $BR(\bar B\rightarrow X_s\gamma)$ does not agree with 
the experimental result have been rejected. We have used $|V_{ts}|=0.04$ 
and $\hat F_{B_s}=238$ MeV but the limits for other values of these 
parameters can be obtained by simple rescalings.  
Horizontal lines in figs. \ref{fig:sfcnc1}a-\ref{fig:sfcnc1}f show the
upper bound (\ref{eqn:Cs_lim}) on 
$(1/2)\left(|\tilde C^\mu_S|^2+|\tilde C^\mu_P|^2\right)$ for 
$\mu_b=4.2$ GeV and 
$\hat F_{B_s}=238$ MeV (solid lines), $\hat F_{B_s}=207$ MeV (dashed lines) 
and $\hat F_{B_s}=269$ MeV (dotted lines). Vertical lines show the
corresponding constraint imposed by the experimental lower limit 
$\Delta M_s>14/$ps (excluded are the points to the left of these lines).

 From figs. \ref{fig:sfcnc1}a-\ref{fig:sfcnc1}f it is clear that 
the lowest possible values of $\hat F_{B_s}$ \cite{LEBE}, 
which in the model independent analysis of the preceding section gave the  
biggest effects of the scalar operators in the 
$\bar B\rightarrow X_s\mu^+\mu^-$ and $\bar B\rightarrow K\mu^+\mu^-$
transitions, are in the MSSM allowed only for small values of the Wilson
coefficients $|C^{\mu(1)}_S|^2$ and $|C^{\mu(1)}_P|^2$. Moreover, for
$M_A>200$ GeV and $\hat F_{B_s}\simgt238$ MeV the lower limit
$\Delta M_s>14/$ps becomes more constraining than the bound 
(\ref{eqn:mainlimit}). This means 
that in the MSSM (or any other model in which ${\cal O}_{S,P}$ arise from 
the FV couplings similar to $\left[X_{\rm LR}\right]^{qb}$ in eq. 
(\ref{eqn:XLR})) the possible effects of the scalar operators 
${\cal O}_{S,P}$ in the 
$\bar B\rightarrow X_s\mu^+\mu^-$ and $\bar B\rightarrow K\mu^+\mu^-$
decays must be smaller than the estimates given in section 
3. For example, using the formulae of section 3 and the numbers that
can be extracted from fig. \ref{fig:sfcnc1}b, we find that for $M_A=300$ 
GeV and $\tan\beta=50$ the maximal effects in the inclusive process are 
bounded by
\begin{eqnarray}
\Delta BR(B\rightarrow X_s\mu^+\mu^-)_{\rm nonres}
\simlt2.2\times10^{-7}
\label{eqn:deltaBRfullmax}
\end{eqnarray}
obtained for $f\approx12/14$ (for which (\ref{eqn:mainlimit}) and
$\Delta M_s>14$/ps allow for maximal value of 
$|\tilde C^\mu_S|\approx|\tilde C^\mu_P|$) and  $r\approx1$, that is, 
any effects of the scalar operators must be below $5$\%. 
The maximal effects in the exclusive decay
$\bar B\rightarrow K\mu^+\mu^-$ are then also suppressed by the
limit on the $B^0_s$-$\bar B^0_s$ mass difference:
\begin{eqnarray}
\Delta BR(B\rightarrow K\mu^+\mu^-)_{\rm nonres}\simlt1.0\times10^{-7}~.
\end{eqnarray}
The suppression further with decreasing value of $\tan\beta$ and 
increasing mass scale of the Higgs boson sector (set by $M_A$) up to 
$M_A\simgt650$ GeV.

Since the effects of the FV couplings (\ref{eqn:FVcouplings}) in
$BR(B^0\rightarrow\mu^+\mu^-)$ scale as $(1/M_A)^4$ while in $\Delta M_s$
only as $(1/M_A)^2$, for sufficiently heavy Higgs sector and sufficiently
large couplings $\left[X_{\rm LR}\right]^{sb}$ it is possible to get from
the formula (\ref{eqn:ddms}) $\delta(\Delta M_s)<2 (\Delta M_s)^{\rm SM}$ 
that is $|\Delta M_s|^{\rm MSSM}$ again compatible with the experimental
lower limit (this possibility is seen in the upper branch of points in figure 
\ref{fig:sfcnc1}f) and, at the same time, $BR(B^0\rightarrow\mu^+\mu^-)$ 
below the CDF upper limit. This happens only
for $M_A\simgt750$ GeV. For such Higgs boson masses and values of 
the couplings $\left[X_{\rm LR}\right]^{sb}$ 
the upper bound (\ref{eqn:Cs_lim}) can be saturated and simultaneously
$\hat F_{B_s}$ can assume lowest possible values obtained from lattice
simulations \cite{LEBE}. Only then could the effects of the scalar 
operators ${\cal O}_{S,P}$ in $\bar B\rightarrow X_s\mu^+\mu^-$ and 
$\bar B\rightarrow K\mu^+\mu^-$ decays reach the maximal values discussed 
in section 3 (reduced only slightly by the fact that in the MSSM 
$r=r^\prime=1$). One should stress, however, that, at least in the
MFV supersymmetry, the couplings 
$\left[X_{\rm LR}\right]^{sb}$ of the required magnitude can be generated
by the chargino stop loops only for very large values of the stop
mixing parameter $A_t$ along with significantly split stop masses
and are very unlikely from the point of view of generation the soft
SUSY breaking terms and most likely leading to the dangerous color
breaking minima of the scalar fields potential.

\section{Conclusions}

Rare decays of $B$ mesons are one of the places where the ongoing 
experimental measurements can reveal effects of new physics. The 
processes involving the $b\rightarrow sl^+l^-$ and $b\rightarrow dl^+l^-$
transitions are particularly interesting in this context. The most
general low energy Hamiltonian describing their phenomenology involves 
the so-called scalar operators ${\cal O}_S^l=(\bar s_Lb_R)(\bar ll)$, 
${\cal O}_P^l=(\bar s_Lb_R)(\bar l\gamma^5l)$ (and similar ones with 
$s_L\rightarrow d_L$). Their Wilson coefficients are negligible in the 
SM but in models of new physics can be quite substantial compared to the 
coefficients of the other, usually studied, operators. This is so for 
example in the minimal supersymmetric extension of the SM even if the 
supersymmetric partners of the known particles are rather heavy, provided 
the ratio of the vacuum expectation values $v_u/v_d=\tan\beta$ of the two 
Higgs doublets is large and the mass scale of the extended Higgs sector 
is not too high. 

In section 3 of this paper, following the earlier work \cite{BOEWKRUR},
we have used the experimental upper limits on the branching fractions
$BR(B^0_{s,d}\rightarrow l^+l^-)$ to place the constraints on the 
Wilson coefficients of the scalar operators relevant for the 
$b\rightarrow sl^+l^-$ and $b\rightarrow dl^+l^-$ transitions. 
Particularly stringent constraint obtained from the  limit 
$BR(B^0_s\rightarrow\mu^+\mu^-)<0.95\times10^{-6}$ has been subsequently 
used to asses in a model independent way the impact the scalar operators
$(\bar s_Lb_R)(\bar\mu\mu)$, $(\bar s_Lb_R)(\bar\mu\gamma^5\mu)$ may have 
on the rates of the inclusive $\bar B\rightarrow X_s\mu^+\mu^-$ and 
exclusive $\bar B\rightarrow K\mu^+\mu^-$ decays. 

We have found that the increase of $BR(\bar B\rightarrow X_s\mu^+\mu^-)$ 
due to the scalar operators cannot exceed (5-15)\% (depending on the
range of the dimuon invariant mass), that is, it is always
smaller than the uncertainty of SM NNLO result which we have estimated 
in section 2. The large effects of the scalar operators found in this 
decay in ref. \cite{WAAT} are therefore already excluded.
On the other hand, the maximal increase of exclusive rate 
$BR(\bar B\rightarrow K\mu^+\mu^-)$ can be still quite large, of order 
$1.7\times10^{-7}$, comparable with the present error of the experimental
result. The latter, when compared to the SM prediction 
$BR(\bar B\rightarrow K\mu^+\mu^-)=(3.5\pm1.2)\times10^{-7}$ 
\cite{ALLUGRHI}, leaves some room for positive new physics contribution.
However the SM prediction for this rate hinges on the theoretical problems 
related to the determination of the relevant nonperturbative formfactors. 
Before this issue is settled
(and the experimental errors shrink) no firm conclusion about the
detectability of new physics effects in the exclusive decay 
$\bar B\rightarrow K\mu^+\mu^-$ can be drawn.

In the supersymmetric scenario with large $\tan\beta$ and not too heavy 
Higgs sector, in which large values of the Wilson coefficients of the 
scalar operators can be naturally generated, the potential effects of 
${\cal O}_S^\mu$ and ${\cal O}_P^\mu$ in $b\rightarrow s\mu^+\mu^-$ are 
further constrained by the experimental lower limit on the 
$B^0_s$-$\bar B^0_s$ mass difference. This has been illustrated in section 
4  in the case of the minimal flavour violation scenario considered in 
papers \cite{BUCHROSL1,BUCHROSL2,BUCHROSL3}. However, the limits
on the scalar operator contributions to $BR(B^0_s\rightarrow \mu^+\mu^-)$,
$BR(\bar B\rightarrow X_s\mu^+\mu^-)$ and
$BR(\bar B\rightarrow K\mu^+\mu^-)$ that can be derived by inserting in 
the formulae of section 3 numbers exctracted from figure \ref{fig:sfcnc1} 
(for different values of $\tan\beta$ and $M_A$) are valid also if the 
flavour violation originates in the squark sector, provided supersymmetric
particles are heavy enough in order not to contribute appreciably to the 
box and vector boson penguin amplitudes. This is because the 
specific relation between the Wilson coefficients of 
${\cal O}_S^\mu$ and ${\cal O}_P^\mu$ and the Wilson coefficients of
the scalar operators contributing to the $B^0_s$-$\bar B^0_s$ mixing 
amplitude relies only on the existence in the low energy effective 
theory of the flavour violating couplings (\ref{eqn:FVcouplings}) and 
not on the specific mechanism of the flavour violation in the underlying
theory.

\vskip0.3cm
\noindent{\sl Note added} While completing this paper we have learned about 
a similar independent study by F. Kr\"uger et al. \cite{KR}. In particular 
they confirm our conclusion that the large effects of the scalar operators
found in ref. \cite{WAAT} in the inclusive rate are already excluded by 
the experimental data.

\vskip0.3cm
\noindent{\bf Acknowledgments} P.H.Ch. would like to thank M. Misiak
for many fruitful discussions related to the subject of this work and
to E. Lunghi and F. Kr\"uger for useful communications. The work
of P.H.Ch. was supported by the EC contract HPRN-CT-2000-0148 and the Polish 
State Committee for Scientific Research grant 2 P03B 129 24 for 2003-2005. 
The work of \L .S. was supported by the Polish State Committee for 
Scientific Research grant 2 P03B 040 24 for 2003-2005.

\end{document}